\documentstyle{mn}

\input{psfig.sty}
\input{epsf.sty}
\def\plotone#1{\centering \leavevmode
\epsfxsize=\columnwidth \epsfbox{#1}}

\def\msun{{\rm M_{\odot}}}

\def\be{\begin{equation}}
\def\ee{\end{equation}}
\def\mdot{\dot M}
\def\te{T_{\rm eff}}
\def\msunyr{{\rm M_\odot yr^{-1}}}

\begin{document}

\title{Mass estimates in short--period compact binaries}
\author[U. Kolb, A. R. King and I. Baraffe]{U. Kolb$^1$,
A. R. King$^2$ and I. Baraffe$^{3}$\\ 
$^1$ Department of Physics \& Astronomy, The Open University, Walton Hall,
Milton Keynes, MK7~6AA (u.c.kolb@open.ac.uk)\\
$^2$ Department of Physics \&Astronomy, University of Leicester,
Leicester, LE1~7RH (ark@star.le.ac.uk)\\ 
$^3$ Ecole Normale Sup\'{e}rieure de Lyon, C.R.A.L.\ (UMR 5574 CNRS), 
F-69364 Lyon Cedex 07, France (ibaraffe@ens-lyon.fr)
}
\date{MNRAS, in press (September 2000)}

\maketitle
\begin{abstract}
Using stellar models we
investigate the relation between mass and the spectral type of the
secondary star in low--mass short--period compact binaries such as
cataclysmic variables and soft X--ray transients. Allowing for
different mass transfer rates and different system ages 
prior to mass transfer we find that the secondaries should populate a
band in the spectral type vs.\ mass plane. 
The mass $M_{\rm ms}$ of a ZAMS star with the same spectral type as
the donor is effectively an upper limit to the donor mass.
The lower mass limit for a given spectral type depends on the mixing
length parameter. If this is large, there is no 
lower limit if the spectral type is later than K6. The band width
decreases from $0.4 \msun$ at K6 to less than $0.2 \msun$ at K0. If the
mixing length parameter is small, there is no lower mass limit if
the spectral type is later than M2, and the band width decreases from 
$0.2 \msun$ at M2 to less than $0.1 \msun$ for types earlier than K0.

We also point out an error in the method suggested by Beekman et al.\
(1997) to estimate the primary mass in a soft X--ray transient. 
\end{abstract}


\section{Introduction}
\label{sec:intro}

Reliable mass estimates for the stellar components of
compact binaries, i.e.\ cataclysmic variables (CVs) and X--ray binaries
(XBs), are notoriously difficult to obtain. Except for the few
double--lined eclipsing systems a successful mass determination
hinges on an estimate of the orbital inclination and
on subtleties in defining the effective dynamical centre of the emission
from the donor star (see e.g.\ Smith \& Dhillon 1998). 

In practice, masses in short--period CVs or XBs are often estimated
with the assumption that the donor star (the secondary) obeys a
main--sequence mass--radius relation. A simple consequence of Roche
geometry and Kepler's law is that in a semidetached binary the orbital
period $P$ and the mean density of the donor are related via   
\begin{equation}
P_h = k_2 \left( \frac{r_2^3}{m_2} \right)^{1/2} 
\label{eq1}
\end{equation}
(e.g.\ King 1988). 
Here $P_h$ denotes the period in hr, $r_2$ and $m_2$ the radius $R_2$
and mass $M_2$ of the donor star in solar units. The coefficient $k_2$
is a weak function of mass ratio $q=M_2/M_1$ ($M_1$ is the primary
mass), with $k_2 \simeq 8.86$ for $q\la1$. If the secondary is in the
core hydrogen burning phase its radius can be expressed as  
\begin{equation}
r_2 = r \, m_2^\alpha,
\label{eq2}
\end{equation}
usually with $\alpha \simeq 1$, $r\simeq 1$, so that (\ref{eq1}) reads  
\begin{equation}
P_h \simeq 9 \, m_2.
\label{eq3}
\end{equation}

In short--period `j--driven' compact binaries, where the orbital angular
momentum losses drive mass transfer, the donor star can deviate from 
this standard mass--radius relation mainly  
for the two following reasons.
 \\  
(i)~The secondary experiences continuous mass loss which
   perturbs thermal equilibrium. The surface luminosity is
   no longer in balance with the luminosity generated in the core by
   nuclear reactions, and the difference causes the star to expand or
   contract. 
   Mass--losing main--sequence stars are smaller than in thermal
   equilibrium when they are predominantly radiative, i.e.\ for 
   $m_2 \ga 0.6$, while they are oversized if they are
   predominantly convective, i.e.\ for $m_2 \la 0.6$ (e.g.\ Whyte \&
   Eggleton 1983; Stehle et al.\ 1996; Baraffe \& Kolb 2000). \\
(ii)~The donor star could have evolved off the zero--age main--sequence
   (ZAMS) prior to 
   mass transfer. Once mass transfer is under way the nuclear
   state at the onset of mass transfer is effectively frozen in. Once
   sufficiently low--mass, these donors mimic stars that are older
   than a Hubble time.  
   Recently, Baraffe \& Kolb (2000; see also Kolb \& Baraffe 2000)
   argued that the observed late spectral type of CV secondaries provide
   strong evidence that a large fraction of CV secondaries have evolved off
   the ZAMS prior to mass transfer. Similarly, theoretical arguments
   suggest that the majority of short--period neutron--star XBs form
   with fairly massive donors ($m_2 \ga 1$), allowing them to age
   prior to mass loss. The constant $r$ in (\ref{eq2}) can therefore vary
   significantly.

Applying (\ref{eq3}) to estimate the donor mass would usually give
a value which is larger than the real mass. The donor is {\em
undermassive} compared to a ZAMS star that fills the Roche lobe in 
the given binary.

In the following we show that the effective temperature of the donor
star --- and hence its spectral type --- is much less sensitive 
to the effects of mass loss and nuclear ageing than the radius. 
Using the latest generation of low--mass star models we 
explore theoretically to what extent the spectral type provides a
reasonable estimate of the donor mass. 
In addition, we reconsider Beekman et al.'s (1997) attempt to place a
lower limit on the black hole (or neutron star) mass in soft X--ray
transients.

\section{The stellar code}

For our numerical experiments we employ the latest generation of
low--mass star and brown dwarf models by Baraffe et al.\ (1995, 1997,
1998, henceforth summarized as BCAH models) which represent  
a significant improvement in the quantitative description of 
stars with mass $\la 1$ $\msun$. 

The main strengths of the models are in two areas: the microphysics 
determining the equation of state in the stellar interior, and
the non--grey atmosphere models which enter as the outer boundary
condition. The equation of state (Saumon, Chabrier and Van Horn 1995)
is specifically calculated for very low--mass stars, brown dwarfs and 
giant planets. Recent much improved cool atmosphere models (Hauschildt
et al.\ 1999; see also the review of Allard et al.\ 1997) now provide
realistic atmosphere profiles, which we use as the outer boundary
condition, and synthetic spectra. Chabrier \& Baraffe (1997) have
shown that evolutionary models employing a grey 
atmosphere instead, e.g.\ the standard Eddington approximation,
overestimate the effective temperature for a given mass, and
yield too large a minimum hydrogen burning mass. 

For a more detailed description of the code and the numerous tests
against observations we refer to Baraffe \& Kolb (2000),
Kolb \& Baraffe (1999; 2000), and references therein.  

We obtain the spectral type of a stellar model from its calculated
colour $(I-K)$ and the empirical SpT$- (I-K)$ relation established
by Beuermann et al.\ (1998). 

One of the main uncertainty in stellar models is the treatment of
convection. Here we use the standard mixing length theory, with the
ratio $\alpha$ = $l_{\rm mix} \, / \, H_{\rm P}$ of  
mixing length $l_{\rm mix}$ and local pressure scale height $H_{\rm
P}$ as a free parameter.
Except for the specific case of the Sun,
$\alpha$ cannot be calibrated very well by a comparison between
observations and models. 
The chief reason for this is that the age and chemical
composition of observed stars is usually not known independently (see 
Baraffe et al.\ 1998; Baraffe 1999).  
According to such comparisons, the most likely range for $\alpha$ 
is between 1 and 2. Recent 2--dim.\ hydrodynamical simulations (Ludwig,
Freytag \& Steffen 1999) also show a variation of 
$\alpha$ with effective temperature $\te$ and surface gravity
$g$ between typically 1 and 2. We therefore adopt
this range of values to characterize the present uncertainty
due to convection.
The spectral type of chemically inhomogeneous stars ($m \ga 0.6 \msun$)
depends on the choice of $\alpha$, while unevolved, fully convective
stellar models are insensitive to $\alpha$.

\section{The donor mass--spectral type relation}

\begin{figure}
\plotone{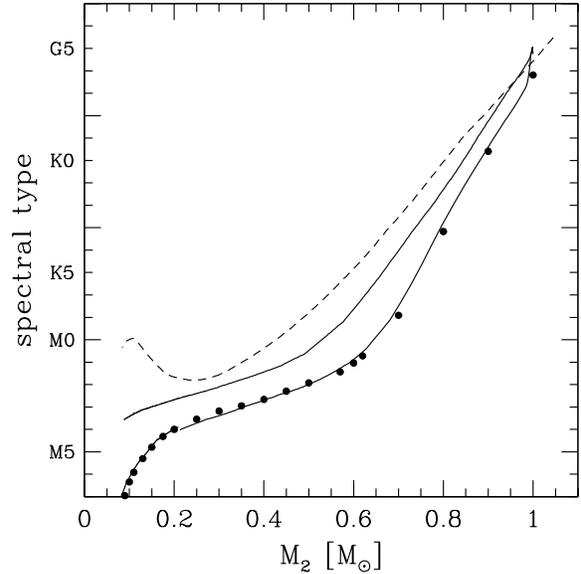}
\caption{
Spectral type versus secondary mass: dependence on the evolutionary state
of the secondary at turn--on of mass transfer. Dots denote ZAMS models.
Solid curve, close to the dots: donor unevolved (standard sequence). 
Solid curve, above the dots: donor moderately evolved
($X_c=0.16$; initial donor mass $M_{2,i}=1.2 \, \msun$). Dashed: donor
close to the terminal main sequence ($X_c= 4 \times 10^{-4}$,
$M_{2,i}=1.0 \, \msun$). All sequences calculated with low mass
transfer rate ($1.5 \times 10^{-9}$ $\msunyr$).
\label{fig:spm1}}
\end{figure}

\begin{figure}
\plotone{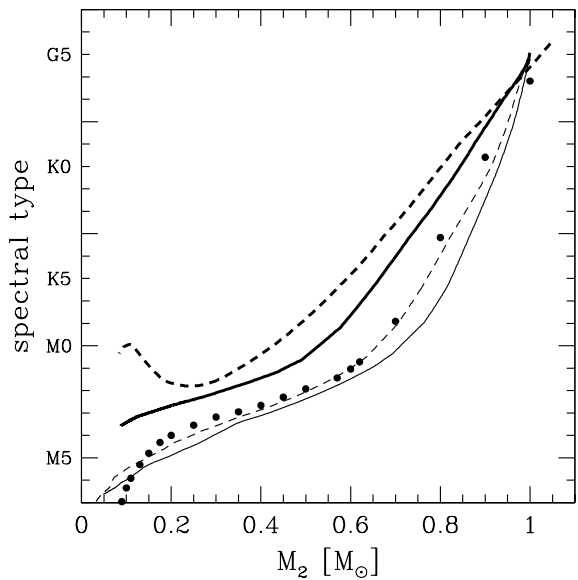}
\caption{
Spectral type versus secondary mass: dependence on the mass transfer
rate. Dots denote ZAMS models. 
Thick lines: evolved donor ($X_c= 4 \times 10^{-4}$); dashed: $\dot M
= 1.5 \times 10^{-9}$ $\msunyr$; solid: $\dot M = 5 \times 10^{-8}$
$\msunyr$. Thin lines: unevolved donors; dashed: $\dot M
= 10^{-8}$ $\msunyr$; solid: $\dot M = 10^{-7}$
$\msunyr$.
\label{fig:spm2}}
\end{figure}

\begin{table}
\caption{Spectral type and corresponding donor mass $M_2$ along the
sequences displayed in Fig.~\protect{\ref{fig:spm3}}. The mass in the
second column (taken from the sequence with initial donor mass 
$M_{2,i}=1.2$ $\msun$, $X_{\rm c} = 4 \times 10^{-4}$, $\dot M = 1.5
\times 10^{-9} \msunyr$, $\alpha=2$) defines an approximate lower limit to the
mass of any donor with spectral type SpT, the mass in the third column  
(from the unevolved sequence with $M_{2,i}=1.0$ $\msun$, 
$\dot M = 10^{-8} \msunyr$, $\alpha=1$) an approximate upper limit. There is no
lower limit for SpT later than K6. 
\label{tab0}
}
\begin{tabular}{ccc}
  SpT & $M_2/\msun$ & $M_2/\msun$ \\
      & lower limit & upper limit \\
   G5 & 0.83  & 1.00 \\
   G6 & 0.79  & 0.98 \\
   G7 & 0.75  & 0.97 \\
   G8 & 0.72  & 0.95 \\
   G9 & 0.68  & 0.93 \\
 &&\\
   K0 & 0.65 & 0.91\\
   K1 & 0.61 & 0.88\\
   K2 & 0.56 & 0.85\\
   K3 & 0.52 & 0.83\\
   K4 & 0.47 & 0.80\\
   K5 & 0.42 & 0.77\\
   K6 & 0.35 & 0.74\\
   K7 & -  & 0.70\\
 &&\\
   M0 & - & 0.66 \\
   M1 & - & 0.60 \\
   M2 & - & 0.50 \\
   M3 & - & 0.38 \\
   M4 & - & 0.24 \\
\end{tabular}
\end{table}

The effective temperature $\te$ of low--mass main--sequence stars is
rather insensitive to their degree of thermal disequilibrium. They behave 
just like giant stars on the Hayashi line,
with nearly constant effective temperature.
This is well understood in terms of the analytic Hayashi line theory
for fully convective stars (e.g.\ Kippenhahn \& Weigert 1990).
Similarly, $\te$ is not very sensitive to the nuclear age of the
star. Single--star tracks in the HR diagram run almost vertical while 
on the lower main--sequence.   

Using BCAH models we quantify the deviation from equilibrium spectral
type, as a measure of the surface temperature, for the donor stars in
j--driven compact binaries with various evolutionary histories. 
We simplify the problem by considering sequences with constant mass
transfer rate $\dot M$. To bracket the evolution of real systems where
$\dot M$ presumably varies with orbital period we calculate sequences
with a high transfer rate and with a low transfer rate.
Test sequences where we computed the transfer rate self--consistently
according to simple orbital angular momentum loss descriptions give
indeed the same qualitative and quantitative results.

We first discuss sequences calculated with solar metallicity
and $\alpha = 1$ (Figs.~\ref{fig:spm1} and \ref{fig:spm2}).

In particular, in Fig.~\ref{fig:spm1} we show the {\em standard
sequence} of Baraffe \& Kolb (2000). Here mass transfer starts from an
unevolved (ZAMS) donor with 
mass 1 $\msun$, proceeds at a constant rate $1.5\times 10^{-9}$
$\msunyr$, stops when the donor becomes fully convective (at mass 0.21
$\msun$), and resumes at the lower rate $5\times 10^{-11}$ $\msunyr$
once the donor has settled back into thermal equilibrium. This sequence
fits the period gap in the framework of disrupted orbital braking
(e.g.\ King 1988). In addition, in Fig.~\ref{fig:spm2} we plot two
other sequences that start from a $1 \, \msun$ ZAMS star (thin lines;
henceforth ``unevolved sequences''), one 
calculated with $\dot M = 10^{-8}$ $\msunyr$, the
other with $10^{-7}$ $\msunyr$. The mass transfer
rates in j--driven compact binaries are 
usually estimated to be well $\la 10^{-8}$ $\msunyr$ (see e.g.\ Warner  
1995, his Fig.~9.8, for CVs; Chen et al.\ 1997 for XBs). A white
dwarf system with $\dot M = 10^{-7}$ $\msunyr$ would not appear as a
CV at all, but presumably as a supersoft X--ray binary (e.g.\ van den
Heuvel 1992; see also King et al.\ 2000).

Fig.~\ref{fig:spm1} shows that the effect of thermal disequilibrium is
negligible along the standard sequence; this hardly
differs from that for systems with a ZAMS donor (i.e.\ in thermal
equilibrium). If $\mdot$ is higher 
than in the standard sequence, the spectral type is slightly later for
a given mass (Fig.~\ref{fig:spm2}). 
Even the extreme sequence with $\dot M = 10^{-7}$ $\msunyr$ never
deviates by more than $0.1 \msun$ for a given SpT from this relation
(Fig.~\ref{fig:spm2}, thin solid line). In other words, unevolved
donor stars follow an 
almost unique SpT--$M_2$ relation, the one given by the ZAMS. 

This relation is much less well--defined when we include systems where
the secondary has already burned a significant fraction of its 
hydrogen supply in the detached phase prior to mass transfer. 
Note that along these ``evolved sequences'' the donor is still in
the core hydrogen burning phase, otherwise the system would evolve off
to longer periods (see.g.\ Savonije \& Pylyser 1988). 

Standard common envelope arguments predict that CVs form over the
whole range of secondary masses, $0.1 \la m_2 \la 1$, hence the
majority of CVs should 
have essentially unevolved donors (Politano 1996, de~Kool 1992).
Yet the generally late spectral type in systems with $P_h \ga  5$
suggests that a fair fraction of CVs actually contain an evolved
secondary (Baraffe \& Kolb 2000; Kolb \& Baraffe 2000; see also Ritter
2000). Similarly, evolutionary considerations suggest that most
short--period neutron--star XBs form with a somewhat evolved
main--sequence donor with mass $\ga 1$ $\msun$ (King \& Kolb 1997;
Kalogera \& Webbink 1998; Kalogera et al.\ 1998). This is mainly
because systems with a more massive secondary stand a greater chance
of surviving the supernova explosion and associated mass ejection at
neutron star formation.

It is therefore important to consider evolved sequences.
Fig.~\ref{fig:spm1} shows --- in addition to the standard sequence
discussed above --- a moderately evolved low $\mdot$ sequence
(initial central hydrogen content $X_c=0.16$, initial donor mass
$M_{2,i}=1.2$ $\msun$, $\mdot = 1.5 \times 10^{-9}$ $\msunyr$) and a
significantly evolved, low $\mdot$ sequence ($X_c=4\times10^{-4}$,
$M_{2,i}=1.2$ $\msun$, $\mdot = 1.5 \times 10^{-9}$ $\msunyr$) where
the donor was almost at the terminal main sequence at turn--on of mass
transfer. 

From Fig.~\ref{fig:spm2} we see that the mass transfer rate has the same
relative effect on evolved and unevolved sequences: the evolutionary
track is shifted vertically downwards, to later spectral types. The
tracks in heavy linestyle are evolved sequences, one with
high mass transfer rate (solid line; $\mdot = 5 \times 10^{-8}$
$\msunyr$; $X_c=4\times10^{-4}$, $M_{2,i}=1.0$ $\msun$), the other
with low mass transfer rate (dashed; $\mdot = 1.5 \times 10^{-9}$
$\msunyr$, $X_c=4\times10^{-4}$, $M_{2,i}=1.2$ $\msun$; same as in
Fig.~\ref{fig:spm1}). 

The highly evolved track with low transfer rate ($\dot M = 1.5 \times
10^{-9} \msunyr$) and the unevolved 
track with high transfer rate ($\dot M = 10^{-8} \msunyr$) should
bracket the location of compact binary secondaries.

\begin{figure}
\plotone{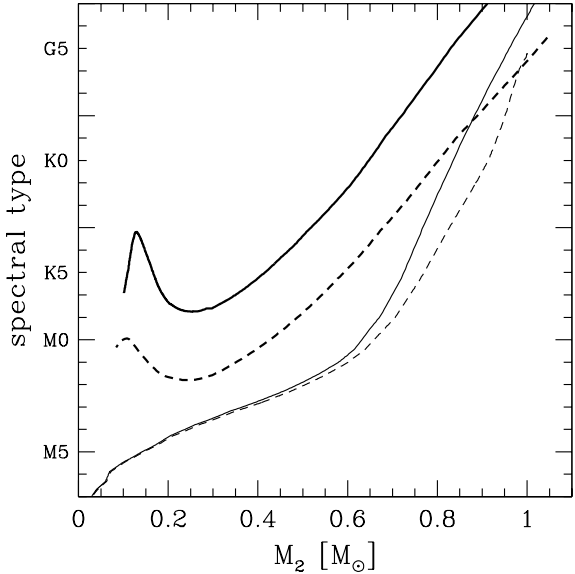}
\caption{Expected populated domain in the spectral type versus
secondary mass plane. The band defined by the unbroken curves   
correspond to $\alpha = 1.9$, the band defined by the dashed curves to 
$\alpha = 1$. (Thick linestyle: evolved sequences with 
$X_c= 4 \times 10^{-4}$, $M_{2,i}=1.2 \, \msun$, $\dot M = 1.5
\times 10^{-9} \msunyr$; thin linestyle: unevolved sequences
with $\dot M = 10^{-8} \msunyr$.)  
\label{fig:spm3}}
\end{figure}

Fig.~\ref{fig:spm3} shows the effect of the mixing length parameter
$\alpha$ on these limiting curves in the SpT vs.\ mass diagram. Unevolved
tracks become undistinguishable for predominantly 
convective donors ($M_2 \la 0.6 \msun$), while the most evolved
sequences are quite sensitive to $\alpha$. Table~\ref{tab0} gives
lower and upper limits to $M_2$ for a given spectral type. The range
reflected in this table is to some extent due to our ignorance of the
appropriate mixing length parameter.

The metallicity $Z$ of the donor can affect the mass--spectral type
relationship as well. Lower $Z$ gives a lower 
mass for a given spectral type. However, as stressed by Beuermann et
al.\ (1998), there is no evidence for low metallicity secondaries
in the currently observed sample of CVs. Hence we restrict
our analysis to systems with solar metallicity ($Z=0.02$).

The observations compiled by Smith \& Dhillon (1998), when taken at
face value, are broadly consistent with the band shown in
Fig.~\ref{fig:spm3} (see their Fig.~3). They do not, however, 
help to constrain the mixing length parameter.
Four out of 22 systems in their sample fall outside the region defined
by Tab.~\ref{tab0}, even when the quoted errorbars are taken into
account: U Gem, IP Peg, CN Ori and A0620-00. The donor masses of two
of these systems have recently been revised to much smaller
values. For U Gem (M4--M5), Long \& Gilliland (1999) give $M_2 \simeq 0.41
\msun$ as preferred value. Given the uncertainty on the
inclination we consider the remaining disagreement with the limits in
Tab.~\ref{tab0} as too marginal to deduce any conflict with
theory. Similarly, Beekman et 
al.\ 2000 find the new value $M_2 \simeq 0.33^{+0.14}_{-0.05} \msun$
for the donor mass of IP Peg. Taking into account the range of
spectral type determinations (M2--M5; Ritter \& Kolb 1998;
Smith \& Dhillon 1998) this system is now in agreement with
Tab.~\ref{tab0}. The mass estimate for the discrepant system CN Ori is
suspect as it is based on radial velocity curves exhibiting
significant phase shifts (Smith \& Dhillon 1998). Finally, the donor
star in A0620-00 clearly is very undermassive and must be 
highly evolved. As the accretor is a black hole the initial donor mass
could have been much larger than $1.5 \msun$ (the value considered here and 
representative for CVs and neutron--star LMXBs). 

If the observed sample is restricted to double--lined, eclipsing CVs
where the secondary mass can be determined independent of the
main--sequence assumption (\ref{eq3}) for the secondary, 
the limits in Tab.~\ref{tab0} are acceptable in all but two cases (data from
Ritter \& Kolb, 1998, updated). The two cases where the agreement is
only marginal are U Gem (see above) and EM Cyg, where the recently
re--determined donor mass, $0.99 \pm 0.12 \msun$ (North et al.\ 2000),
appears to be somewhat too large for the spectral type K3.

\section{Limits on the primary mass}

This work originally started as an attempt to improve on a 
a paper by Beekman et al.\ (1997: hereafter B97) who considered
limits to the primary mass in soft X--ray transients. Although it
turned out that our results do not 
significantly improve these limits, we wish to comment on
the method 
suggested by B97 as a whole, and point out an error  
in their analysis, which so far has not been corrected in the
literature. 

B97 made use of an inequality derived by King, Kolb  
\& Burderi 1996 (hereafter KKB), which expresses the condition for a
short--period low--mass X--ray binary to have an unstable accretion
disc and thus appear as a transient. Unfortunately B97 misunderstood the
definition of the quantity $\hat 
m_2$ in this inequality. $\hat m_2$ is defined by KKB as the ratio of
the secondary mass $m_2$ to that of the ZAMS star, $m_{\rm ms}(\rho)$  
(in solar units), which would fill its Roche lobe in a binary of the
same orbital period,  
\begin{equation}
\hat m_2 \equiv {m_2\over m_{\rm ms}(\rho)}.
                               \label{4}
\end{equation}
As is obvious from (\ref{eq1}) the relevant main--sequence star has
the same {\it mean density} as the secondary; hence the notation
$m_{\rm ms}(\rho)$. B97 confused $\hat m_2$ with the ratio $m_2/m_{\rm
ms}(R)$, where $m_{\rm ms}(R)$ is the mass (in solar units) of a ZAMS
star that has the same {\it absolute radius} $R_2$ as the
secondary. It can be shown that  
\begin{equation}
m_{\rm ms}(R) = {m_2\over \hat m_2^{2/3}},
\label{12}
\end{equation}
so that a lower bound on $\hat m_2$ (which is required
to obtain a primary mass limit) is  
\begin{equation}
\hat m_2 > \biggl[{\rho_2\over \rho_{\rm ms}(T_{\rm eff})}\biggr]^{3/2}.
                               \label{3}
\end{equation}
Here $\rho_2$ is the mean density of the secondary, (given directly by
the binary period, see Eq.~(\ref{eq1}), and $\rho_{\rm ms}(T_{\rm
eff})$ is the mean density of the main--sequence star with the same
spectral type as the secondary. This is weaker by a factor
$[\rho_2/\rho_{\rm ms}(T_{\rm eff})]^{1/2}$ compared to B97's limit,
which in turn weakens the limits on $M_1$.

In addition to this obvious mistake, we caution
that the limitations and uncertainties inherent in using the disc
instability criterion are probably too severe to give a reliable mass
limit. This criterion assumes that the binary  
transfers mass as a result of angular momentum losses (`j--driven')
rather than nuclear expansion of the secondary (`n--driven'), but
allows for the fact that this star may be somewhat nuclear--evolved
before mass transfer begins. It is therefore only appropriate for
short orbital periods.   
Although the disc instability criterion correctly represents
the statistical properties of the observed sample of soft X--ray
transients, its 
application to individual systems is suspect for at least two reasons. 

1.~The criterion is based on a comparison of the secular mean
  mass transfer rate driven by magnetic braking and the critical mass
  transfer rate for disc stability. Hence uncertainties in 
  the magnetic braking strength, in particular at long orbital
  periods, and in the critical rate (see KKB; King, Kolb
  \& Szuszkiewicz 1997; Dubus et al.\ 1999) propagate directly into the
  mass limit. 

2.~It is not clear if a given system is j--driven, in
  particular if the period is $\ga 10$~h. The ultimate fate of a
  system as j-- or n--driven is determined by its orbital period $P_i$
  when mass transfer begins. If this exceeds a certain `bifurcation
  period' $P_{\rm bif}$ (Pylyser \& Savonije, 1988, 1989) the binary
  eventually becomes a diverging n--driven system, while if $P_i <
  P_{\rm bif}$ it ends up as j--driven. Uncertainties in for example
  the strength of magnetic braking at periods $\ga 12$ hr make $P_{\rm
  bif}$ correspondingly uncertain. Further, even given a definite
  value of $P_{\rm bif}$, n--driven systems born near this period
  initially evolve to periods below it (Pylyser \& Savonije, 1988,
  1989) before nuclear evolution eventually wins out and drives the
  system to long periods. 
  This comment is particularly relevant for the neutron--star transients 
  Cen X--4 and Aql X--1.

\section{Summary}

We have applied up--to--date stellar models of low--mass main--sequence
stars to investigate the mass--spectral type relation if these stars
experience mass loss and are somewhat nuclear--evolved. 

Unevolved stars follow an almost unique SpT--$M_2$ relation, whatever 
the mass transfer rate. Nuclear--evolved stars have a somewhat earlier
spectral type than a ZAMS star with the same mass. The spectral type
is earlier if the star is more evolved. 

Accurate mass predictions for a given spectral type are hampered by the
present uncertainty of the mixing length parameter. 
The mass $M_{\rm ms}$ of a ZAMS star with the same spectral type as
the donor is effectively an upper limit to the donor mass.
The lower mass limit for a given spectral type depends on the mixing
length parameter. If this is large, there is no 
lower limit if the spectral type is later than K6. The band width
decreases from $0.4 \msun$ at K6 to less than $0.2 \msun$ at K0. If the
mixing length parameter is small, there is no lower mass limit if
the spectral type is later than M2, and the band width decreases from 
$0.2 \msun$ at M2 to less than $0.1 \msun$ for types earlier than K0.

A logical next step is to consider the evolution of thermally unstable
systems that start mass transfer from a fairly massive donor ($M_2 \ga
2 \msun$) and reappear as j--driven short--period systems
after reversal of the mass ratio. Detailed calculations of such
systems are underway (Schenker et al.\ 
2000, in preparation). These will allow us to assess how different
descendants of thermal--timescale mass transfer are from the
systems considered here, which had a more moderate initial mass ($<1.5
\msun$) and were j--driven from the start. 

The results presented here are rather robust and inevitable
predictions of standard stellar structure and evolution theory
that has been successfully tested against observations of isolated
low--mass stars. It is therefore highly desirable to test them by
observational means, i.e.\ by a reliable donor mass determination in
systems with spectral type in the indicated range.

\section*{Acknowledgments}

We thank Hans Ritter and Klaus Schenker for useful discussions.
ARK thanks the U.K.\ Particle Physics and Astronomy 
Research Council for a Senior Fellowship. 
Comments by the anonymous
referee helped to improve this paper.
We thank EGIDE (ALLIANCE contract 000193RL)
for travel and visiting support.


\end{document}